\documentclass{rs_main}

\usepackage[normalem]{ulem}
\usepackage{siunitx}
\usepackage{url}
\usepackage{xr-hyper}
\makeatletter
\newcommand*{\addFileDependency}[1]{%
  \typeout{(#1)}\@addtofilelist{#1}%
  \IfFileExists{#1}{}{\typeout{No file #1.}}}
\makeatother
\Mtitle{Diagnosing parameter unidentifiability in stochastic dynamics with an application to flocking models}
\Mauthor{Natsuki Kurita$^{1}$ and Hiroshi Kori$^{1, 2}$}
\Maffil{
    $^{1}$ \textit{Department of Mathematical Informatics, Graduate School of Information Science and Technology, The University of Tokyo, Tokyo, Japan}\\
$^{2}$ \textit{Department of Complexity Sciences and Engineering, Graduate School of Frontier Sciences, The University of Tokyo, Chiba, Japan}
}
\Mcorres{Natsuki Kurita (ORCID: 0009-0007-1298-6777)\\
\email{natsuki.kurita.research@gmail.com}}
\Msubject{mathematical biology, applied mathematics, computational biology}
\Mkeywords{parameter estimation, parameter identifiability, stochastic process, flocking, collective motion}

\begin{document}

\makeMtitle

\begin{abstract}
    Although flocking behaviour has been studied extensively in active matter physics and mathematical biology, methods for assessing parameter estimability remain underdeveloped. For models formulated as stochastic differential equations (SDEs), we develop a Hessian-based diagnostic framework that detects unidentifiable directions and indicates whether, and how, estimation precision can be improved. The framework utilizes the submartingale structure of the loss function's Hessian and tracks its minimum eigenvalue over time. Under convergence of the time-averaged Hessian, we prove that this eigenvalue grows linearly for identifiable models, but only sublinearly, as $o(t)$, for unidentifiable ones; for a specific type of parameter redundancy, this $o(t)$ bound is tightened to $o(\sqrt{t}\log t)$. Applying the framework to stochastic flocking models, we demonstrate that, in identifiable settings, the magnitude of the eigenvalue reflects the attainable estimation precision. We further uncover a physical mechanism of unidentifiability: as the system approaches the ordered state, the parameters become algebraically degenerate, even though the model contains no trivial redundancy. Identifiability therefore depends on the dynamical regime, not only on model structure. Our framework offers a practical tool for reliable parameter estimation, not only in flocking models but in SDE models across diverse fields.
\end{abstract}

\section{Introduction}
Flocking behaviour, a ubiquitous natural phenomenon characterized by the coordinated motion of biological groups, has been an active area of research in mathematical biology and active matter physics~\cite{Vicsek2012, Marchetti2013, Sumpter2005}.
Since Reynolds~\cite{Reynolds1987} proposed a pioneering paradigm to reproduce collective motion in computer animation, numerous models have been introduced to uncover the underlying mechanisms of group dynamics. Among them, the seminal Vicsek model~\cite{Vicsek1995} framed collective motion within the context of non-equilibrium statistical mechanics. More recently, to capture stochastic fluctuations in the behaviour of individual biological agents, models formulated as stochastic differential equations (SDEs) have been gaining prominence~\cite{Gautrais2009, Gautrais2012, Takahashi2020, Klamser2021}.

In parallel, advances in tracking technology have produced large volumes of individual-level trajectory data for bird flocks and fish schools~\cite{Ballerini2008, Nagy2010}, enabling interaction rules to be inferred directly from data~\cite{Lukeman2010, Katz2011, Herbert-Read2011, Heras2019}. 
However, these studies typically extract interaction rules non-parametrically or fit only a subset of model parameters, while forward simulation with pre-assigned parameters still plays a central role in model-based studies. Consequently, it remains challenging to rigorously confront mechanistic models with empirical trajectories by estimating all parameters and assessing the reliability of those estimates. 

One of the few exceptions is the work by Gautrais et al.~\cite{Gautrais2012}, who constructed an SDE flocking model based on empirical trajectory data of \textit{Kuhlia mugil} and estimated its parameters via nonlinear regression. While effective, such nonlinear regression methods struggle to assess whether the estimated parameters are reliable in principle. Typically, estimation precision is inferred \textit{a posteriori} by evaluating the empirical variance of the estimates across repeated trials. This approach is useful, but does not reveal the underlying causes of such variation, nor does it provide clear guidelines on whether and how the precision can be improved. In another notable study involving parameter fitting, Klamser and Romanczuk~\cite{Klamser2021} applied evolutionary algorithms to investigate prey grouping under predation pressure. However, several parameters were fixed to pre-assigned values, thereby reducing the number of parameters to be estimated. Such simplifications highlight the difficulty of estimating all parameters simultaneously.

To overcome these challenges in stochastic models, quasi-maximum likelihood estimation (QMLE)~\cite{Florens-zmirou1989, Kessler1997, Sorensen2004} offers a promising approach. QMLE maximizes an approximate likelihood function derived from the Euler--Maruyama discretization of SDEs. In addition to providing an estimate of its precision via the Hessian matrix of the loss function, it is particularly suitable for two reasons. First, its asymptotic properties, including consistency and asymptotic normality, are supported by a well-established theory under standard regularity conditions~\cite{Yoshida1992, Uchida2012}. Second, empirical trajectory data naturally correspond to discretely observed time series of the SDE variables and can thus be used directly with minimal preprocessing.

\begin{figure}[tbhp]
        \centering
        \includegraphics[width=\textwidth]{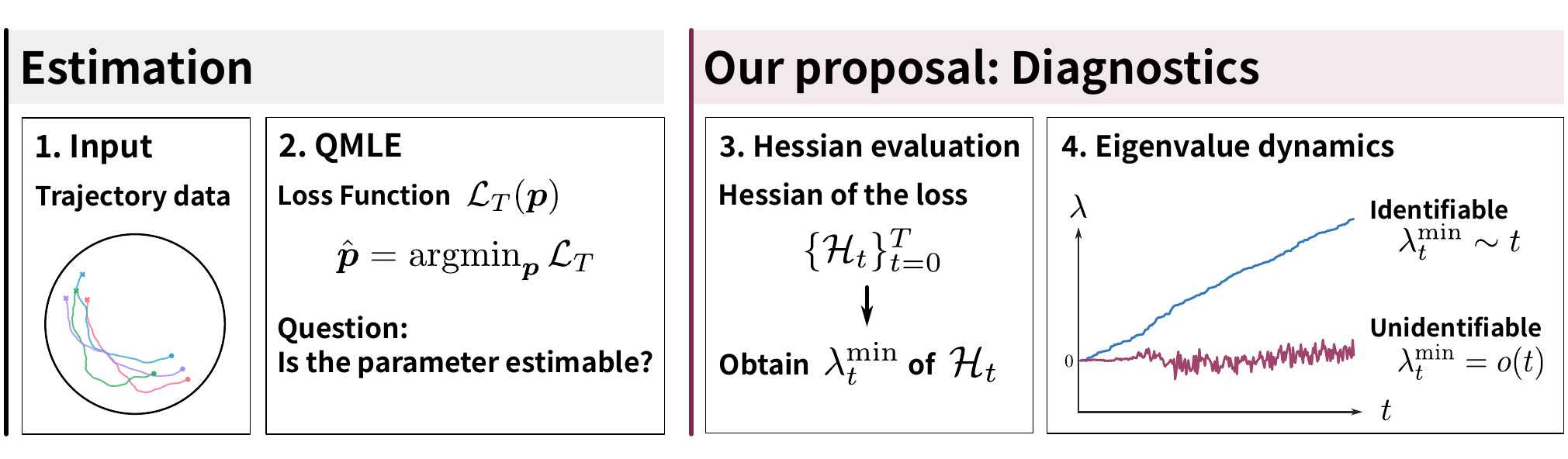}
        \caption{Overview of the proposed framework and typical minimum-eigenvalue dynamics. We propose a diagnostic procedure attached to QMLE that detects parameter unidentifiability in SDE models by monitoring the minimum eigenvalue of the loss Hessian.}
        \label{fig:overview}
\end{figure}

Even though this estimation method is highly effective, a fundamental challenge remains: parameter unidentifiability, which arises when changes in one parameter can be compensated by changes in others~\cite{Raue2009}. A closely related phenomenon, termed sloppiness~\cite{Brown2003, Gutenkunst2007, Transtrum2015}, has been extensively studied in ordinary differential equation (ODE) models, where the eigenvalue spectrum of the Hessian matrix characterizes parameter identifiability. Moreover, diagnostic procedures employing the profile likelihood have been successfully used to detect structural and practical unidentifiability in ODE models~\cite{Raue2009, Kreutz2013}. To our knowledge, however, few diagnostics are tailored to the temporal structure of SDE models. To address this gap, we establish a mathematical framework by applying martingale theory to the Hessian matrix of the loss function and propose a practical diagnostic based on it. We show that parameter identifiability is reflected not in a static snapshot of the minimum eigenvalue but in its stochastic temporal behaviour. Specifically, we demonstrate that the minimum eigenvalue of identifiable models grows linearly, so that noise becomes relatively negligible, while for unidentifiable models it grows sublinearly as $o(t)$ and is dominated by fluctuations (Figure~\ref{fig:overview}).

To demonstrate the practical utility of our framework, we apply it to flocking models. By systematically combining two dynamical frameworks with four representative interaction rules~\cite{Vicsek1995, Gautrais2012, Takahashi2020, Klamser2021} and a null baseline, we construct and analyse ten model variants. Consistent with our theory, we observe two distinct scaling regimes in the eigenvalue time series: linear growth and sublinear fluctuation. These reveal that four of the ten models, namely those involving the Klamser--Romanczuk (KR2021) and Takahashi--Komeyama (TK2020, for two-agent groups) social interactions, exhibit parameter unidentifiability. For the identifiable models, we further show that the estimation error decreases in proportion to the inverse square root of the minimum eigenvalue, which is analogous to the familiar $1/\sqrt{n}$ law of classical statistics under asymptotic normality. The minimum eigenvalue thus serves as an effective sample size that quantifies attainable precision. Finally, we trace the unidentifiability to compensatory relationships between parameters and show that, for KR2021, this degeneracy arises near the limiting ordered state even though the model contains no explicitly redundant parameters.

Together, these results constitute a first step towards a data-driven approach to flocking research, in which mechanistic models are rigorously confronted with empirical trajectories through the estimation of all parameters and the assessment of their reliability. More broadly, our framework applies to a wide class of SDE models whose parameters are estimated from discretely observed data, and thus offers a general tool for diagnosing parameter identifiability across diverse fields.

\section{Estimation framework and flocking models}\label{sec:conventional}
    \subsection{Quasi-maximum likelihood estimation}
    \label{sec:qmle}
    QMLE is a variation of maximum likelihood estimation (MLE) whose loss function is derived via the Euler--Maruyama discretization of SDEs. To ensure broad applicability of this procedure beyond flocking models, we consider a system of SDEs that covers a wide variety of specific formulations:
    \begin{equation}\label{eq:continuous_sde}
            dX_i = a_i(\boldsymbol{X}(t), \boldsymbol{Y}(t), \boldsymbol{p})dt + \sigma_i(\boldsymbol{X}(t), \boldsymbol{Y}(t), \boldsymbol{p}) dW_i \quad (i=1,2,\ldots,N), 
    \end{equation}
    where $X_i\in\mathbb{R}$ is the state variable of the $i$-th agent and $\boldsymbol{X} := (X_1,\dots,X_N)^\top$ is the collective state vector. In the flocking models examined in this study, $X_i$ represents the heading angle $\theta_i$ or the angular velocity $\omega_i$. The functions $a_i$ and $\sigma_i$ are the drift and the diffusion coefficients parameterized by $\boldsymbol{p}\in \mathbb{R}^{d_p}$, respectively. The $W_i$ are mutually independent standard Wiener processes. The vector $\boldsymbol{Y}\in\mathbb{R}^{d_a}$ is an auxiliary state vector that contains quantities such as the agent positions and their pairwise distances. Its value at time $t$ is uniquely determined by its initial value $\boldsymbol{Y}(0)$ and by the history of $\boldsymbol{X}$ up to time $t$. Applying the Euler--Maruyama discretization to Equation~\eqref{eq:continuous_sde}, we obtain the discretized SDE:
    \begin{equation}\label{eq:discrete_sde}
            X_{i,n} - X_{i,n-1} = a_i(\boldsymbol{X}_{n-1}, \boldsymbol{Y}_{n-1},\boldsymbol{p}) \Delta t + \sigma_i(\boldsymbol{X}_{n-1}, \boldsymbol{Y}_{n-1}, \boldsymbol{p}) \Delta W_{i, n},
    \end{equation}
    where $\Delta t$ is the time step and $\Delta W_{i,n} $ are mutually independent Wiener increments that are also independent of $\{\boldsymbol{X}_m, \boldsymbol{Y}_m\}_{m=0}^{n-1}$. Hereafter, $t$ denotes the step index rather than physical time. Since every Wiener increment obeys $\mathcal{N}(0, \Delta t)$, the transition probability density of the original continuous SDE is approximated by a Gaussian distribution $\mathcal{N}(X_{i,n-1}+a_{i,n-1}\Delta t,\ \sigma_{i,n-1}^2\Delta t)$. Taking the negative logarithm of the joint transition density over $t$ time steps, we obtain the loss function $\mathcal{L}_t$. By introducing simplified notation $a_{i,n}(\boldsymbol{p}) := a_i(\boldsymbol{X}_{n}, \boldsymbol{Y}_n, \boldsymbol{p})$, $\sigma_{i,n}(\boldsymbol{p}) := \sigma_i(\boldsymbol{X}_{n}, \boldsymbol{Y}_n, \boldsymbol{p})$, and $F_{i,n}(\boldsymbol{p}) := X_{i,n} - X_{i,n-1} - a_{i,n-1}(\boldsymbol{p}) \Delta t$ for $n \geq 1$, the loss function $\mathcal{L}_{t}$ and the estimator are written as
    \begin{align}
    \label{eq:contrast_function}
        \mathcal{L}_t(\boldsymbol{p}) &= \sum_{n=1}^t\sum_{i=1}^N\left[ \frac{1}{2}\log(2\pi \Delta t) + \log(\sigma_{i,n-1}) + \frac{F_{i,n}^2}{2\sigma_{i,n-1}^2 \Delta t}\right], \\
        \label{eq:qmle}
        \hat{\boldsymbol{p}} &:= \operatorname*{argmin}_{\boldsymbol{p}}\mathcal{L}_T,
    \end{align}
    where $T$ denotes the final step. The diffusion coefficients $\sigma_i$ are assumed to be strictly positive, as is the case for all models in Section~\ref{sec:flocking_models}. This estimator is interpreted as an approximate maximum likelihood estimator, and its consistency in the high-frequency regime has been established under standard regularity conditions~\cite{Yoshida1992, Uchida2012}. 

    Note that the loss function is structurally similar to the one typically used for parameter estimation in ODE systems with observation noise~\cite{Villaverde2022, Philipps2025}. However, a fundamental distinction lies in the underlying noise models. QMLE accounts for intrinsic process noise driven by the Wiener terms within the system dynamics. In contrast, standard ODE estimation methods model measurement noise, mathematically represented as a sequence of independent Gaussian errors superimposed onto deterministic ODE trajectories.

    \subsection{Flocking models}
    \label{sec:flocking_models}
    This study employs ten models of collective motion constructed by combining dynamical frameworks and social interaction rules taken from preceding studies~\cite{Vicsek1995, Kuramoto1975, Gautrais2012, Klamser2021, Takahashi2020}. All models operate under the assumption of a constant speed $v_0$.

    The first dynamical framework, referred to as \textbf{overdamped orientational dynamics}, describes the time evolution of the heading angle $\theta_i$:
    \begin{equation}\label{eq:model_theta}
        d\theta_i = \omega_i^{\mathrm{soc}}dt + \sigma_\theta dW_i,
    \end{equation} 
    where $\omega_i^{\mathrm{soc}}$ denotes the angular velocity induced by social interaction, and the parameter $\sigma_\theta$ controls the intensity of the angular noise. For a point particle moving at constant speed, Newton's equation reduces to this form: a lateral force $F_\perp$ changes the heading at a rate $F_\perp/(M v_0)$, where $M$ is the mass, without rotational inertia.
    
    The second framework, referred to as \textbf{inertial orientational dynamics}, introduces orientational inertia. This formulation was first proposed as the persistent turning walker model by Gautrais et al.~\cite{Gautrais2012}, and can be viewed as an Ornstein--Uhlenbeck (OU) process with a time-varying target:
    \begin{equation}\label{eq:model_omega}
        d\omega_i = -\frac{\omega_i - \omega_i^{\mathrm{soc}}}{\tau}dt + \sigma_\omega dW_i,
    \end{equation}
    where $\tau$ is the relaxation time over which $\omega_i$ approaches the target turning rate $\omega_i^{\mathrm{soc}}$, and $\sigma_\omega$ denotes the noise intensity of the angular velocity. The heading angle is obtained by integrating $\omega_i$ over time, i.e., $d\theta_i/dt = \omega_i$.

    The concrete functional forms of $\omega_i^{\mathrm{soc}}$ and the ground-truth parameters are specified in Table~\ref{tab:summary_of_modules} and ~\ref{tab:setting_01_1_parameters}, and detailed descriptions of the original models are provided in the Supplementary Information, Section 3. The social interaction term $\omega_i^{\mathrm{soc}}$ depends on the basic state variables, such as position, velocity, and heading angle of the focal agent and its neighbours; neighbours are defined by the Voronoi tessellation~\cite{Vandenheuvel2024}. The Voronoi neighbourhood captures metric-free interactions and visual occlusion effects common in biological collectives such as bird flocks~\cite{Ballerini2008, Ginelli2010} and fish schools~\cite{Strandburg-Peshkin2013}.
    Because the social interaction $\omega_i^{\mathrm{soc}}$ is defined independently of the dynamical framework, the two components can be combined in a modular manner: two dynamical frameworks and five social interaction terms (including the null interaction) yield $2\times5=10$ model variants.

    \begin{table}[htbp]
        \centering
        \caption{Social interaction formulations. $\mathcal{N}_i$ is the set of Voronoi neighbours of agent $i$; $\boldsymbol{e}_i^\perp$ is the unit vector obtained by rotating $\boldsymbol{v}_i/\|\boldsymbol{v}_i\|$ by $\pi/2$ anticlockwise; $\phi_{ij}$ is the bearing angle of agent $j$ relative to the heading of agent $i$; $\bar{\boldsymbol{x}}$ and $\bar{\boldsymbol{v}}$ are the mean position and velocity of the neighbours of agent $i$ (excluding $i$ itself); $d^{\mathrm{eq}}$ is the equilibrium distance; and $m$ is a steepness parameter. Superscripts indicate the role of each coupling, and accents distinguish parameters with different physical dimensions.}
        \label{tab:summary_of_modules}
        \renewcommand{\arraystretch}{2.5} 
        \begin{tabular}{llc}
            \toprule
            Module name & Equation & Reference \\
            \midrule 
            Null & 0 & Baseline \\
            KV & $ \dfrac{1}{|\mathcal{N}_i|}\displaystyle\sum_{j\in\mathcal{N}_i}k^{\mathrm{alg}}\sin(\theta_j-\theta_i)$ &~\cite{Vicsek1995, Kuramoto1975} \\
            G2012 & $ \dfrac{1}{|\mathcal{N}_i|}\displaystyle\sum_{j\in \mathcal{N}_i} \left[ k^{\mathrm{alg}} \sin(\theta_j - \theta_i) + \tilde{k}^{\mathrm{pos}} \|\boldsymbol{x}_j-\boldsymbol{x}_i\| \sin\phi_{ij}\right]$ &~\cite{Gautrais2012} \\
            KR2021 & $\dfrac{1}{|\mathcal{N}_i|}\displaystyle\sum_{j\in \mathcal{N}_i} \left[ k^{\mathrm{alg}}\sin(\theta_j-\theta_i) + \hat{k}^{\mathrm{pos}} \tanh(m(\|\boldsymbol{x}_j-\boldsymbol{x}_i\|-d^{\mathrm{eq}}))\sin\phi_{ij} \right]$ &~\cite{Klamser2021} \\
            TK2020 & $\left[-\dfrac{1}{|\mathcal{N}_i|}\displaystyle\sum_{j\in\mathcal{N}_i}k^{\mathrm{sep}}\frac{\boldsymbol{x}_j-\boldsymbol{x}_i}{\|\boldsymbol{x}_j-\boldsymbol{x}_i\|} + k^{\mathrm{coh}}\dfrac{\bar{\boldsymbol{x}}- \boldsymbol{x}_i}{\|\bar{\boldsymbol{x}}- \boldsymbol{x}_i\|} + k^{\mathrm{alg}}\dfrac{\bar{\boldsymbol{v}}- \boldsymbol{v}_i}{\|\bar{\boldsymbol{v}}- \boldsymbol{v}_i\|} \right]\cdot \boldsymbol{e}_i^\perp$ &~\cite{Takahashi2020}\\
            \bottomrule
        \end{tabular}
    \end{table}

    \begin{table}[htb]
    \centering
    \caption{Ground-truth parameters used in the numerical experiments.}
    \label{tab:setting_01_1_parameters}
    \begin{tabular}{llll}
        \toprule
        Category / Model & Parameter & Ground-truth & Unit \\
        \midrule
        \multicolumn{4}{l}{\textbf{Dynamics}} \\
        \qquad \quad Overdamped & $\sigma_\theta$ & 0.5 & $\mathrm{rad\,s^{-1/2}}$ \\
        \qquad \quad Inertial   & $\sigma_\omega$ & 0.5 & $\mathrm{rad\,s^{-3/2}}$ \\
        \qquad \quad Inertial   & $\tau$          & 0.1 & $\mathrm{s}$ \\
        \addlinespace
        \multicolumn{4}{l}{\textbf{Social interactions}} \\
        \qquad \quad KV     & $k^{\mathrm{alg}}$ & 2.0 & $\mathrm{rad\,s^{-1}}$ \\
        \addlinespace
        \qquad \quad G2012  & $k^{\mathrm{alg}}$ & 1.0 & $\mathrm{rad\,s^{-1}}$\\
        \qquad \quad G2012  & $\tilde{k}^{\mathrm{pos}}$ & 1.0 & $\mathrm{rad\,s^{-1}\,m^{-1}}$\\
        \addlinespace
        \qquad \quad KR2021 & $k^{\mathrm{alg}}$ & 1.0 & $\mathrm{rad\, s^{-1}}$ \\
        \qquad \quad KR2021 & $\hat{k}^{\mathrm{pos}}$ & 1.0 & $\mathrm{rad\, s^{-1}}$ \\
        \qquad \quad KR2021 & $m$              & 1.0 & $\mathrm{m^{-1}}$\\
        \qquad \quad KR2021 & $d^{\mathrm{eq}}$  & 0.02 & $\mathrm{m}$ \\
        \addlinespace
        \qquad \quad TK2020 & $k^{\mathrm{sep}}$ & 0.4 & $\mathrm{rad\,s^{-1}}$ \\
        \qquad \quad TK2020 & $k^{\mathrm{coh}}$ & 0.5 & $\mathrm{rad\,s^{-1}}$ \\
        \qquad \quad TK2020 & $k^{\mathrm{alg}}$ & 0.2 & $\mathrm{rad\,s^{-1}}$ \\
        \bottomrule
    \end{tabular}
\end{table}

    \section{Proposed identifiability diagnostic}\label{sec:proposal}
    
    \subsection{Practical procedure}
    \label{sec:practical_procedure}
    Our proposed diagnostic focuses on the temporal dynamics of the minimum Hessian eigenvalue. The core intuition relies on the geometric properties of the loss landscape: the minimum eigenvalue and its corresponding eigenvector identify the flattest direction around the optimal parameter, along which the estimate is least constrained by the data. Assuming that the estimation step has already been completed in the manner of Section~\ref{sec:qmle}, the diagnostic procedure consists of the following steps: 
    \begin{enumerate}[label=\textup{(\roman*)}, align=left, leftmargin=*]
    \item With the parameter fixed at the final estimate $\hat{\boldsymbol{p}}$, evaluate the sequence of log-Hessians $\mathcal{H}_1(\hat{\boldsymbol{q}}), \mathcal{H}_2(\hat{\boldsymbol{q}}), \dots, \mathcal{H}_T(\hat{\boldsymbol{q}})$ (Equation~\eqref{eq:log_hessian}), where $\hat{\boldsymbol q}=\log\hat{\boldsymbol p}$, and compute each minimum eigenvalue $\lambda^{\min}(\mathcal{H}_t(\hat{q}))$.
    \item Plot these minimum eigenvalues against $t=1,\dots,T$ to inspect their temporal scaling.
    \end{enumerate}

    Because stochastic-process theory predicts distinct asymptotic scaling regimes, visual inspection of the minimum-eigenvalue time series provides a practical diagnostic of parameter identifiability:

    \begin{enumerate}[label=\textup{(\alph*)}, align=left, leftmargin=*]
    \item \textbf{Identifiable:}\\
    The minimum eigenvalue exhibits linear growth with minor fluctuations. This implies that the estimation accuracy improves as the data length $T$ increases.
    \item \textbf{Structurally unidentifiable:}\\
       The minimum eigenvalue exhibits sublinear stagnation, typically dominated by noisy fluctuations. In this case, collecting data over a longer period does not resolve the estimation uncertainty, signalling a need for model reformulation when unique parameter identification is critical. 
    \item \textbf{Practically unidentifiable:}\\
    Linear growth with an extremely small slope falls into this class. The minimum eigenvalue increases slowly, or the accumulating signal is obscured by fluctuations visually similar to those in a structurally unidentifiable case. This case requires prohibitively long observations to improve accuracy.
    \end{enumerate}

    This diagnostic has two major purposes. First, the procedure detects inherent parameter unidentifiability prior to full-scale data acquisition. By testing a model against synthetic data generated by the model itself, one can verify whether the true parameters are recoverable under ideal conditions. A persistent fluctuation in the minimum-eigenvalue time series suggests that some parameter combinations may not be reliably identifiable under the assumed observation conditions (see Section~\ref{sec:numerical_verification}). Second, the trajectory of the minimum eigenvalue provides guidance on the observation length required to attain a target level of precision. Although the correspondence has not been proved rigorously, we observe error decay consistent with the growth of the minimum eigenvalue, analogous to the asymptotic relation between Fisher information and variance (see Section~\ref{sec:precision}).
    
    The procedure can also be applied to empirical data because it does not require knowledge of the ground-truth parameters. In this case as well, the reliability of the estimates is supported by the linear growth of the eigenvalues, which corresponds to the progressively deepening minimum in the loss landscape. In this study, however, we focus on the two purposes described above, which require only synthetic data, and demonstrate the practical utility of the procedure in Section~\ref{sec:applications}.

    \subsection{Theoretical guarantees}
    \label{sec:theory}
    The diagnostic procedure described in Section~\ref{sec:practical_procedure} relies on the distinct scaling behaviours of the minimum eigenvalue. To rigorously establish this property, we examine the Hessian of the loss function in detail. Since this section relies on measure-theoretic probability, readers focusing primarily on biological implications and practical procedures may skip this section without loss of continuity.
    
    First, we outline the fundamental assumptions imposed on the SDE systems.
    \begin{condition}[Assumptions on SDEs]\leavevmode
            \label{cond:continuous_sde_assumption}
            \begin{enumerate}[label=\textup{(\roman*)}, align=left, leftmargin=*]
                \item The SDEs are interpreted in the Itô sense.
                \item The parameter space $\mathcal{P}$ is an open subset of $\mathbb{R}_{>0}^{d_p}$. The positivity of parameters ensures that the non-dimensionalized Hessian \textup{(\ref{eq:log_hessian})} is well-defined, and openness is required to support differentiation with respect to parameters.
                \item The diffusion term is non-vanishing, i.e., $\sigma_i(\boldsymbol{X}, \boldsymbol{Y},\boldsymbol{p}) > 0 \quad \text{for all } \boldsymbol{X}\in \mathbb{R}^N, \boldsymbol{Y}\in\mathbb{R}^{d_a}, \boldsymbol{p}\in\mathcal{P}$, and $i=1,2,\ldots,N$.
                \item The drift $a_i$ and the diffusion $\sigma_i$ are twice continuously differentiable ($C^2$) with respect to the parameters. More precisely, for arbitrary states $\boldsymbol{X}$ and $\boldsymbol{Y}$, $a_i(\boldsymbol{X}, \boldsymbol{Y}, \cdot): \mathcal{P} \rightarrow \mathbb{R}$ and $\sigma_i(\boldsymbol{X}, \boldsymbol{Y}, \cdot): \mathcal{P} \rightarrow \mathbb{R}_{>0}$ are $C^2$ functions for all $i=1,2,\ldots,N$.
                \item Every term in the gradient (Supplementary Information, S21) and the Hessian~\eqref{eq:hessian_matrix} is integrable. 
            \end{enumerate}
        \end{condition}
        
    These assumptions provide the foundation for analysing the Hessian. By differentiating the loss function~\eqref{eq:contrast_function} twice with respect to $\boldsymbol{p}$, we obtain the raw Hessian in an explicit form:
    \begin{align}
        \label{eq:hessian_matrix}
        \begin{split}
            \tilde{\mathcal{H}}_t :=  \sum_{n=1}^t \sum_{i=1}^{N} &\left[ \left(-\frac{1}{\sigma_{i,n-1}^2} + \frac{3F_{i,n}^2}{\sigma_{i,n-1}^4\Delta t}\right)(\nabla_{\boldsymbol{p}}\sigma_{i,n-1})(\nabla_{\boldsymbol{p}}\sigma_{i,n-1})^\top \right. \\
                &\quad + \frac{2F_{i,n}}{\sigma_{i,n-1}^3}\left[(\nabla_{\boldsymbol{p}} a_{i,n-1})(\nabla_{\boldsymbol{p}}\sigma_{i,n-1})^\top + (\nabla_{\boldsymbol{p}}\sigma_{i,n-1})(\nabla_{\boldsymbol{p}} a_{i,n-1})^\top\right] \\ 
                &\quad + \left(\frac{1}{\sigma_{i,n-1}} - \frac{F_{i,n}^2}{\sigma_{i,n-1}^3\Delta t}\right)\nabla_{\boldsymbol{p}}^2\sigma_{i,n-1} \\
                &\quad \left. - \frac{F_{i,n}}{\sigma_{i,n-1}^2}\nabla_{\boldsymbol{p}}^2 a_{i,n-1} + \frac{\Delta t}{\sigma_{i,n-1}^2}(\nabla_{\boldsymbol{p}} a_{i,n-1})(\nabla_{\boldsymbol{p}} a_{i,n-1})^\top \right].
        \end{split}
    \end{align}

    The most important structural feature of $\tilde{\mathcal H}_t$ is that it is a matrix-valued submartingale. Let $\mathcal{F}_t$ be the $\sigma$-field generated by $\boldsymbol{X}_0, \boldsymbol{X}_1, \dots,\boldsymbol{X}_t$ and $\boldsymbol{Y}_0$, which intuitively represents all information up to step $t$. By taking the conditional expectation, we obtain the submartingale property of the raw Hessian evaluated at the true parameter $\boldsymbol{p}^*$:
    \begin{equation}
        E[\tilde{\mathcal{H}}_{t+1}(\boldsymbol{p}^*)-\tilde{\mathcal{H}}_{t}(\boldsymbol{p}^*)\mid \mathcal{F}_t] = \sum_{i=1}^N \frac{1}{\sigma_{i,t}^2} \left[2(\nabla_{\boldsymbol{p}}\sigma_{i,t})(\nabla_{\boldsymbol{p}}\sigma_{i,t})^\top
            + \Delta t(\nabla_{\boldsymbol{p}} a_{i,t})(\nabla_{\boldsymbol{p}} a_{i,t})^\top \right]\succeq O,
    \end{equation}
    because the terms linear in $F_{i,t+1}(\boldsymbol{p}^*)$ have zero conditional mean, while the quadratic terms are evaluated using $E[F_{i,t+1}^2(\boldsymbol{p}^*)\mid\mathcal F_{t}]=\Delta t\sigma_{i,t}^2$. The symbol $\succeq$ stands for the positive semi-definite (L\"{o}wner) ordering for symmetric matrices. Intuitively, this inequality reflects the non-decreasing accumulation of information, and hence of estimation precision. Moreover, submartingales admit the Doob decomposition:
    \begin{equation}
    \tilde{\mathcal{H}}_t(\boldsymbol{p}^*) = \tilde{\mathcal{A}}_t(\boldsymbol{p}^*) + \tilde{\mathcal{M}}_t(\boldsymbol{p}^*),
    \end{equation}
    where $\tilde{\mathcal{A}}_t$ and $\tilde{\mathcal{M}}_t$ are the predictable non-decreasing term and the martingale term (Supplementary Lemma S7) of $\tilde{\mathcal{H}}_t$, respectively. The predictable non-decreasing term $\tilde{\mathcal{A}}_t$ represents the accumulation of information over the observation, while $\tilde{\mathcal{M}}_t$ stands for the residual fluctuation. 

    Before analysing the eigenvalues, we must resolve the dimensional mismatch among the elements of the raw Hessian $\tilde{\mathcal{H}}_t$ to make eigenvalue analysis physically consistent. We non-dimensionalize the Hessian following the approach of Gutenkunst et al.~\cite{Gutenkunst2007} via a logarithmic parameter transformation:
    \begin{align}\label{eq:log_hessian}
        \left(\mathcal{H}_t\right)_{ij} &:= \dfrac{\partial^2 \mathcal{L}_t}{\partial \log p_i \partial \log p_j}.
    \end{align}

    This transformation is well-defined due to the positivity assumption on the parameters (Condition~\ref{cond:continuous_sde_assumption}(ii)). If this assumption does not hold, one can instead non-dimensionalize the Hessian by rescaling each parameter with a representative value $p_i^{(0)}$ and differentiating with respect to $p_i/p_i^{(0)}$, at the cost of some arbitrariness in the choice of $p_i^{(0)}$. In what follows, we reserve the notation $\boldsymbol{q}$ for the log-transform of $\boldsymbol{p}$. Note that the log-Hessian~\eqref{eq:log_hessian} is also a submartingale at $\boldsymbol{q}^*$ (Supplementary Theorem S16), and therefore admits the Doob decomposition. We denote the Doob decomposition of the log-Hessian as $\mathcal{H}_t(\boldsymbol{q}^*) = \mathcal{A}_t(\boldsymbol{q}^*) + \mathcal{M}_t(\boldsymbol{q}^*)$ without the tilde. With this notation in place, we obtain the following general scaling result:
    \begin{theorem}[Supplementary Theorem S21]
    \label{thm:asymptotic_growth_rate}
            Let $\Delta t$ be fixed and $\lambda^{\min}(\mathcal{H}_t)$ be the minimum eigenvalue of the log-Hessian $\mathcal{H}_t(\boldsymbol{q}^*)$.
            Additionally, suppose that the time-averaged Hessian $\mathcal{H}_t/t$ converges to a limiting matrix $H_\infty$ almost surely in the Frobenius norm. Then, the minimum eigenvalue is asymptotically given as 
            \begin{equation}
                \lambda^{\min}(\mathcal{H}_t) = \lambda_\infty^{\min}\cdot t + o(t)\ \operatorname{a.s.}
            \end{equation}
            where $\lambda_\infty^{\min}$ is the minimum eigenvalue of $H_\infty$.
    \end{theorem}

    Theorem~\ref{thm:asymptotic_growth_rate} implies a dichotomy in eigenvalue growth rate. If $\lambda_\infty^{\min}>0$, the eigenvalue grows asymptotically linearly, while $\lambda_\infty^{\min}=0$ corresponds to sublinear stagnation characteristic of unidentifiability. Although absolute positions are non-stationary because the agents continue to move, the distribution of relative state variables, such as the heading differences and pairwise distances, may approach a stationary distribution, providing heuristic support for the assumed existence of the limiting matrix.

    Moreover, the scaling can be tightened in the specific case that corresponds to an exact compensatory relationship. A typical cause of unidentifiability is parameter redundancy, whereby different parameter values produce exactly the same dynamics. For example, two parameters may appear only through their product $p_1p_2$, as with an interaction strength multiplied by a scaling factor. We call this severe form of redundancy \textbf{complete degeneracy} (Supplementary Definition S18). In this case, the upper bound of the growth rate is tightened from $o(t)$ to $o(\sqrt{t}\log t)$:
    \begin{theorem}[Supplementary Theorem S22]\label{thm:eigenvalue_envelope_completely_degenerate}
            Let $\Delta t$ be fixed and $\{\mathcal{H}_t(\boldsymbol{q}^*)\}_{t\geq 0}$ be the log-Hessian of a completely degenerate model. Assume in addition that $E[\|\mathcal{M}_{t+1}-\mathcal{M}_t\|_F^2\mid \mathcal{F}_t] \leq C$ for some constant $C>0$ and all $t\geq0$, where $\|\cdot\|_F$ stands for the Frobenius norm. Then, 
            \begin{equation}
            \lim_{t\rightarrow \infty}
                \frac{\lambda^{\min}(\mathcal{H}_t)}{\sqrt{t}\log t} = 0\ \operatorname{a.s.}
            \end{equation}
    \end{theorem}

    One gap between theory and practice is that the practical diagnostic relies on the estimated parameter $\hat{\boldsymbol{p}}$, even though the theorems require the true parameter. This gap remains theoretically open. Nevertheless, the numerical experiments suggest that this substitution preserves the asymptotic scaling behaviour. A notable difference arises for Overdamped-KR2021, where evaluation at $\boldsymbol{p}^*$ yields only slow linear growth except for three samples; we attribute this difference primarily to the location of the evaluation point along the degenerate manifold (see Section~\ref{sec:physical_origin} and Supplementary Information, Section 5.1).

\section{Applications to flocking models}
    \label{sec:applications}
    \subsection{Configuration of the numerical experiments}
    For each of the ten model combinations, we generated 50 independent trajectories for group sizes of $N=2$ and $N=4$ agents. The SDEs were integrated with the Euler--Maruyama scheme using a time step of $\Delta t = 1/24\,\mathrm{s}$, matching a typical \SI{24}{Hz} video frame rate. The same time step was also used in the loss function, so that the simulation, estimation, and diagnostic procedures are mutually consistent. We assumed an open field without boundaries, and each trajectory was simulated for \SI{7}{min}, of which the first \SI{2}{min} were discarded as burn-in to remove the influence of the initial conditions and the remaining \SI{5}{min} were retained for analysis. No artificial measurement noise was added, and complete observation without missing data was assumed.

    For the parameter estimation step, we used the trust-region interior-point method (\texttt{trust-constr}) implemented in SciPy~\cite{Virtanen2020}. We computed the gradient and Hessian by automatic differentiation~\cite{Griewank2008} implemented via JAX~\cite{Jax2018}. Automatic differentiation decomposes a programmed function into a composition of elementary operations and propagates derivatives through it by the chain rule. Unlike finite-difference approximations, it returns derivatives that are exact up to floating-point round-off. This precision is essential here, because the smallest Hessian eigenvalues of interest fall several orders of magnitude below the largest ones and would be obscured by numerical error. The model used for estimation was identical to that used to generate the synthetic trajectories.
    
    For the diagnostic step, we evaluated the log-Hessian matrix at $\boldsymbol{p} = \hat{\boldsymbol{p}}$ for every $t=1,\dots,T$, where $T=\num{7200}$ corresponds to \SI{5}{min} at \SI{24}{Hz}, and analysed its eigenvalues. Results evaluated at the ground truth are given in Supplementary Information, Section 5.1 for comparison.

    The figures below are based on the $N=2$ data; the corresponding results for $N=4$ are qualitatively similar except for TK2020 (see Section~\ref{sec:physical_origin} and Supplementary Information, Section 5.3).

    \subsection{Numerical verification of the predicted scaling law}\label{sec:numerical_verification}
\begin{figure}[htbp]
        \centering
        \begin{subfigure}[b]{0.48\textwidth}
            \textbf{(a)}\vspace{2pt}\\
            \includegraphics[width=\textwidth]{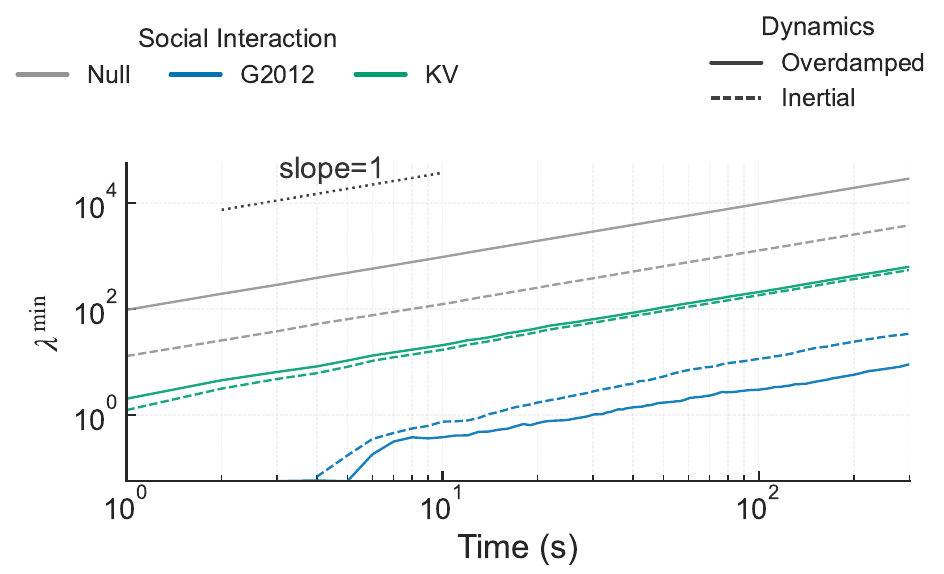}
            \phantomcaption\label{fig:sample_timeseries_identifiable}
        \end{subfigure}
        \hfill
        \begin{subfigure}[b]{0.48\textwidth}
            \textbf{(b)}\vspace{2pt}\\
            \includegraphics[width=\textwidth]{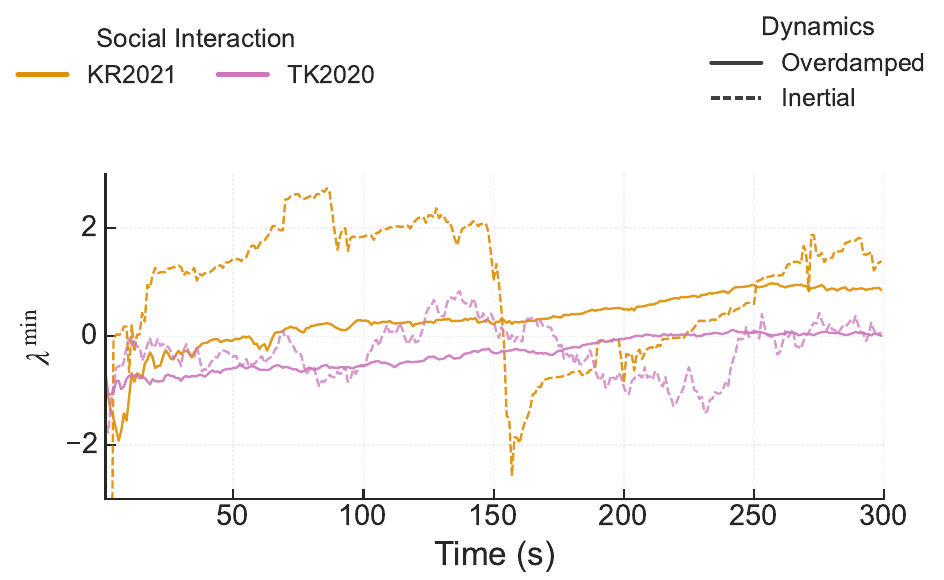}
            \phantomcaption\label{fig:sample_timeseries_unidentifiable}
        \end{subfigure}
        
        \caption{Representative minimum-eigenvalue time series plotted at one-second intervals. 
        \textbf{(a)}~Identifiable cases (log--log scale). The vertical offsets, which reflect the proportionality coefficients, span about four orders of magnitude, as estimated at $t=T$. Negative values at small $t$ are not shown. 
        \textbf{(b)}~Unidentifiable cases (both axes linear). The time series stagnate around zero with fluctuations.}
        \label{fig:sample_time_series}
    \end{figure}

    Figure~\ref{fig:sample_timeseries_identifiable} and Figure~\ref{fig:sample_timeseries_unidentifiable} show representative minimum-eigenvalue time series for models exhibiting linear growth and fluctuation-dominated stagnation, respectively. In Figure~\ref{fig:sample_timeseries_identifiable}, although the eigenvalues can be negative for small $t$, they eventually stabilize and grow in proportion to the observation time. Notably, models with the null interaction fall into this category, implying that the two dynamical frameworks themselves are intrinsically identifiable. The growth rate, however, varies considerably among identifiable models. The proportionality coefficients, estimated from the values at $t=T$, span roughly four orders of magnitude. A excessively small proportionality coefficient can cause practical unidentifiability, as is the case for the KR2021 interaction (discussed in Section~\ref{sec:physical_origin}). In contrast to the identifiable cases, Figure~\ref{fig:sample_timeseries_unidentifiable} illustrates volatile fluctuations characteristic of unidentifiable models. Rather than growing in proportion to time, these time series stagnate around zero with varying volatility depending on the dynamics.

\begin{figure}[htbp]
        \centering
        \begin{subfigure}[b]{0.48\textwidth}
            \textbf{(a)}\vspace{2pt}\\
            \includegraphics[width=\textwidth]{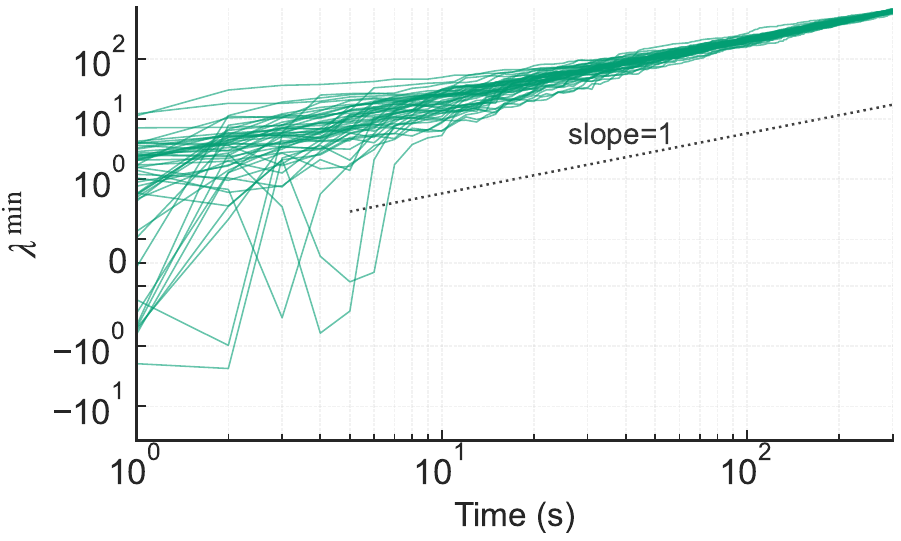}
            \phantomcaption\label{fig:sample_envelope_identifiable}
        \end{subfigure}
        \hfill
        \begin{subfigure}[b]{0.48\textwidth}
            \textbf{(b)}\vspace{2pt}\\
            \includegraphics[width=\textwidth]{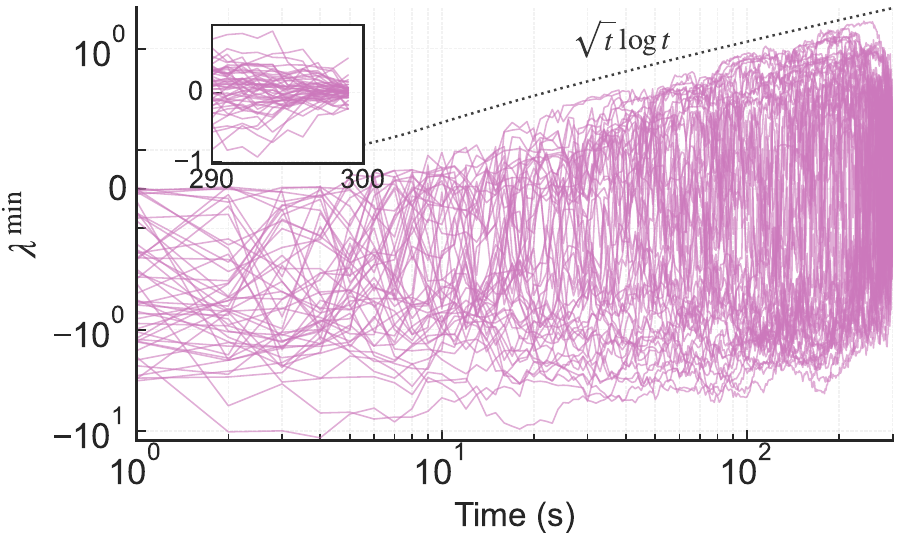}
            \phantomcaption\label{fig:sample_envelope_unidentifiable}
        \end{subfigure}\\
        \begin{subfigure}[b]{0.48\textwidth}
            \textbf{(c)}\vspace{2pt}\\
            \includegraphics[width=\textwidth]{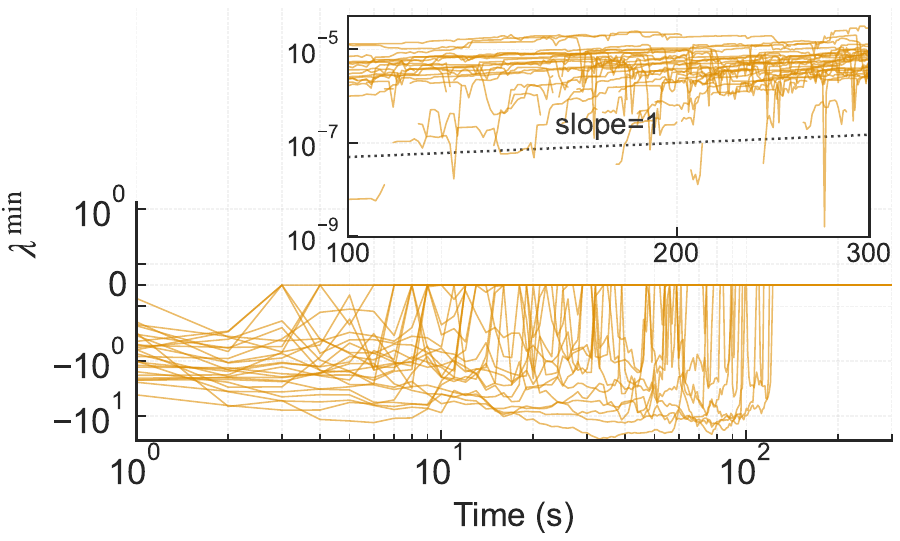}
            \phantomcaption\label{fig:envelope_kr2021_linear}
        \end{subfigure}
        \hfill
        \begin{subfigure}[b]{0.48\textwidth}
            \textbf{(d)}\vspace{2pt}\\
            \includegraphics[width=\textwidth]{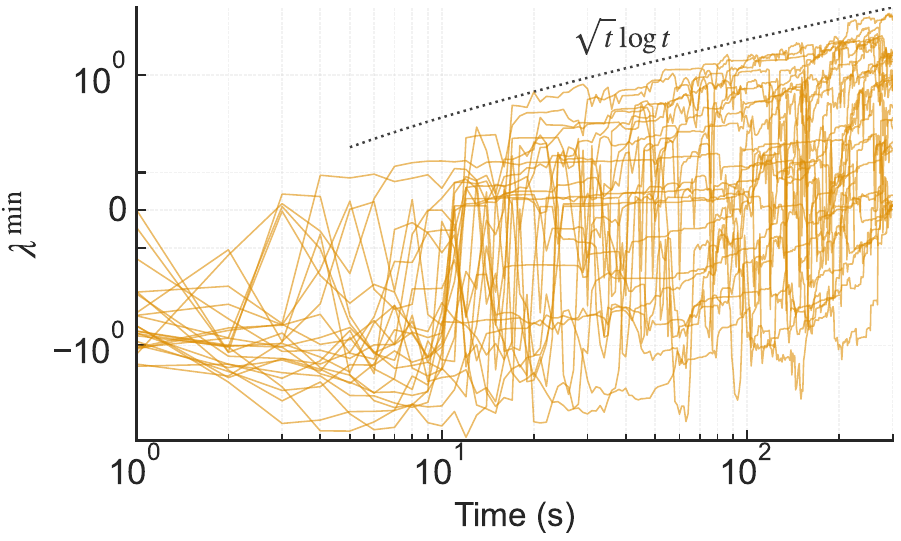}
            \phantomcaption\label{fig:envelope_kr2021_fluctuation}
        \end{subfigure}
        
        \caption{Representative aggregate eigenvalue scaling. Panels (a) and (b) each show 50 sample time series; panels (c) and (d) together show the 50 samples of the Overdamped-KR2021 model, divided according to the estimated value of $m$. The vertical axis uses a symlog scale with a linear region over $[-0.1, 0.1]$. Dashed lines are visual references proportional to $t$ (in (a) and the inset of (c)) and to $\sqrt{t}\log t$ (in (b) and (d)), scaled for visibility.
    \textbf{(a)}~Overdamped-KV model. All samples collapse onto a common linear trend.
    \textbf{(b)}~Overdamped-TK2020 model. The sample paths grow no faster than the $\sqrt{t}\log t$ reference and concentrate near zero at the final time. The inset shows the last \SI{10}{s} on linear axes.
    \textbf{(c), (d)}~Overdamped-KR2021 model. Collectively, both extremely slow linear growth and sublinear fluctuation are observed. The inset in (c) shows the last \SI{200}{s} on log--log axes. Samples in (c) and (d) correspond to the Overdamped-KR2021 estimates in the small-$m$ and large-$m$ regions of Figure~\ref{fig:degeneracy_scatter_KR2021}, respectively.
    }
        \label{fig:envelope}
    \end{figure}

    Figure~\ref{fig:envelope} visualizes the aggregate scaling over the 50 samples for representative overdamped models; the inertial counterparts show qualitatively similar behaviour, except that fluctuation-dominated paths are rarer for Inertial-KR2021 (Supplementary Information, Section 5.2). In identifiable cases, such as the Overdamped-KV model in Figure~\ref{fig:sample_envelope_identifiable}, all sample paths collapse onto a common linear trend. In contrast, for the Overdamped-TK2020 model (Figure~\ref{fig:sample_envelope_unidentifiable}), the paths grow no faster than the $\sqrt{t}\log t$ reference over the observed time scale.

    At the final time step, all sample paths concentrate around zero for the Overdamped-TK2020 model, including slightly negative values. Theoretically, the final value should be non-negative; the slightly negative values observed here are artefacts of the optimization and the log-transformation. Provided that the optimizer attains at least a local minimum in the interior of the parameter space, the gradient vanishes and the Hessian $\tilde{\mathcal{H}}_T$ at the optimum is positive semi-definite. Since the log-Hessian generally satisfies
    \begin{equation}
        \mathcal{H}_t(\boldsymbol{q}) = D(\boldsymbol{p})^\top \tilde{\mathcal{H}}_t(\boldsymbol{p}) D(\boldsymbol{p}) + D(\boldsymbol{p})^\top \operatorname{diag}\bigl(\nabla_{\boldsymbol{p}} \mathcal{L}_t(\boldsymbol{p})\bigr),\quad D(\boldsymbol{p}) = \operatorname{diag}(\boldsymbol{p}),
    \end{equation}
    its final value is positive semi-definite at any local minimum. In practice, however, the optimizer terminates at a finite tolerance, so the gradient $\nabla_{\boldsymbol{p}}\mathcal{L}_T(\hat{\boldsymbol{p}})$ does not vanish exactly, which can leave slightly negative values at the final step. Note that such concentration near zero is not expected at $\boldsymbol{p}^*$, because the loss function does not necessarily attain a local minimum there (see Supplementary Information, Section 5.1).
    
    The KR2021 models exhibit a mixture of identifiable-like and unidentifiable-like behaviour; some samples grow linearly with an extremely small slope while others exhibit fluctuations similar to those of TK2020. As argued in Section~\ref{sec:physical_origin}, the KR2021 models exhibit a mild degeneracy that emerges near the limiting ordered state, and are thus regarded as practically unidentifiable. In such cases, both linear growth and fluctuation-dominated stagnation can occur within the same model.
    
    For the Inertial-KR2021 model, we also observed unexpected behaviour in which some paths decline linearly into the negative region (Supplementary Information, Section 5.2). Geometrically, this corresponds to a saddle that sharpens over time, suggesting that these estimates are trapped at saddle points. 

    \subsection{Impact on estimation precision}\label{sec:precision}
    In statistics, the observed Fisher information, i.e., the Hessian of the negative log-likelihood, and hence its eigenvalues, govern the asymptotic variance of the estimated parameters. Although our Hessian cannot be directly identified with the Fisher information due to the Euler--Maruyama approximation relative to the continuous-time model and the replacement of $\boldsymbol{p}^*$ with $\hat{\boldsymbol{p}}$, our results reproduce analogous tendencies. 
    
\begin{figure}[htbp]
        \centering
        \begin{subfigure}[b]{0.48\textwidth}
            \textbf{(a)}\vspace{2pt}\\
            \includegraphics[width=\textwidth]{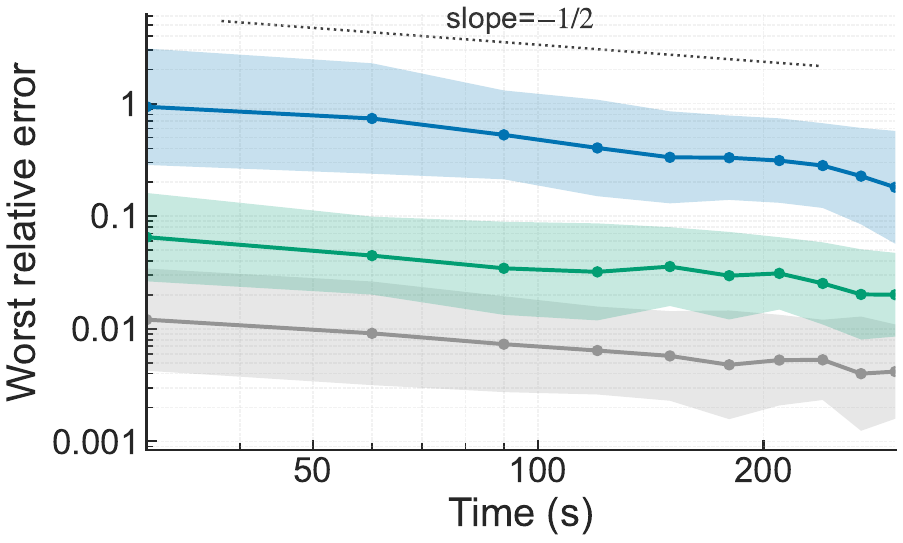}
            \phantomcaption\label{fig:precision_overdamped_identifiable}
        \end{subfigure}
        \hfill
        \begin{subfigure}[b]{0.48\textwidth}
            \textbf{(b)}\vspace{2pt}\\
            \includegraphics[width=\textwidth]{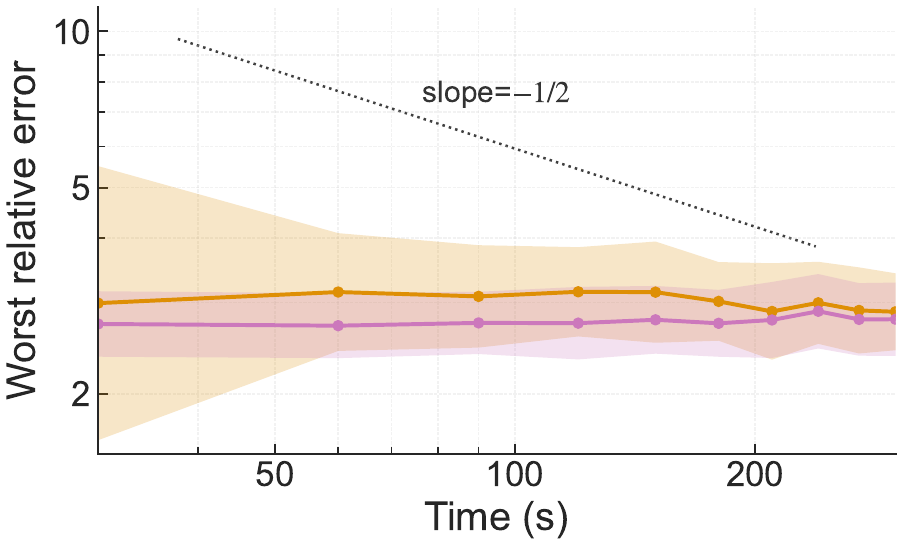}
            \phantomcaption\label{fig:precision_overdamped_unidentifiable}
        \end{subfigure}
        
        \caption{Evolution of the worst relative error over 50 samples for the overdamped models. Each estimate $\hat{\boldsymbol{p}}_t$ is obtained with the data up to $30, 60, \dots, 300$\,s. Lines and bands show the geometric mean and the range of $\pm1$ standard deviation of $\log e_t^{\text{worst}}$ over 50 samples, respectively. \textbf{(a)}~Identifiable cases. The error decreases as observations accumulate, with a slope of approximately $-1/2$ on the log--log scale. 
        \textbf{(b)}~Unidentifiable counterparts. The error does not diminish even with longer observations.}
        \label{fig:precision}
    \end{figure}

    As seen in Figure~\ref{fig:precision_overdamped_identifiable}, the worst relative errors decay with narrowing variance for identifiable models as the data length increases. The worst relative error is defined as
    \begin{equation}
        e_t^{\text{worst}} := \max_{k\in \{1,\dots,d_{p}\}}\left| \log \frac{\hat{p}_{k,t}}{p_k^*} \right| = \max_{k\in \{1,\dots, d_p\}}\left| \hat{q}_{k,t}-q^*_k \right|,
    \end{equation}
    where $\hat{\boldsymbol{p}}_t$ is obtained by re-optimizing the loss function using only the data up to step $t$, in contrast to the final estimate $\hat{\boldsymbol{p}}=\hat{\boldsymbol{p}}_T$ at which the Hessian is evaluated elsewhere in this paper. This criterion treats overestimation and underestimation symmetrically (e.g., a factor of $2$ and a factor of $1/2$ are penalized equally), and reduces to the ordinary relative error $|\hat{p}_k-p_k^*|/p_k^*$ when $\hat{p}_k/p^*_k \approx 1$. Under asymptotic normality $\sqrt{t}(\hat{\boldsymbol{q}}_t-\boldsymbol{q}^*) \xrightarrow{d} \mathcal{N}_{d_p}(\boldsymbol{0}, \mathcal{I}^{-1})$, where $\mathcal{I}$ is the Fisher information per step, each component of the error decays at the rate $1/\sqrt{t}$, as indeed seen in Figure~\ref{fig:precision_overdamped_identifiable}. When the minimum eigenvalue grows proportionally to $t$, the worst error is expected to scale asymptotically as
    \begin{equation}
        e_t^\text{worst}\sim (\lambda^{\min}(\mathcal{H}_t))^{-1/2}.
    \end{equation}
    
    This relation is analogous to the classical result that the standard error of a maximum likelihood estimator scales as the inverse square root of the Fisher information. Here, $\mathcal{H}_t$ plays the role of the accumulated observed Fisher information, and its minimum eigenvalue quantifies the information along the least informative direction in parameter space.
    
    In contrast, unidentifiable models suffer from persistent estimation errors regardless of the data length. In these models, the stagnation of $\lambda^{\min}$ near zero reflects a persistently flat valley in the loss landscape. Parameter estimates can therefore drift along this degenerate direction with little change in the likelihood, which prevents the true parameter from being pinned down.

    \subsection{Physical origin of unidentifiability}
    \label{sec:physical_origin}
     In both TK2020 and KR2021, unidentifiability stems from compensatory relationships among parameters, which can be traced back to the structure of the interaction terms. 

\begin{figure}[tbhp]
        \centering
        \begin{subfigure}[b]{0.48\textwidth}
            \textbf{(a)}\vspace{2pt}\\
            \includegraphics[width=\textwidth]{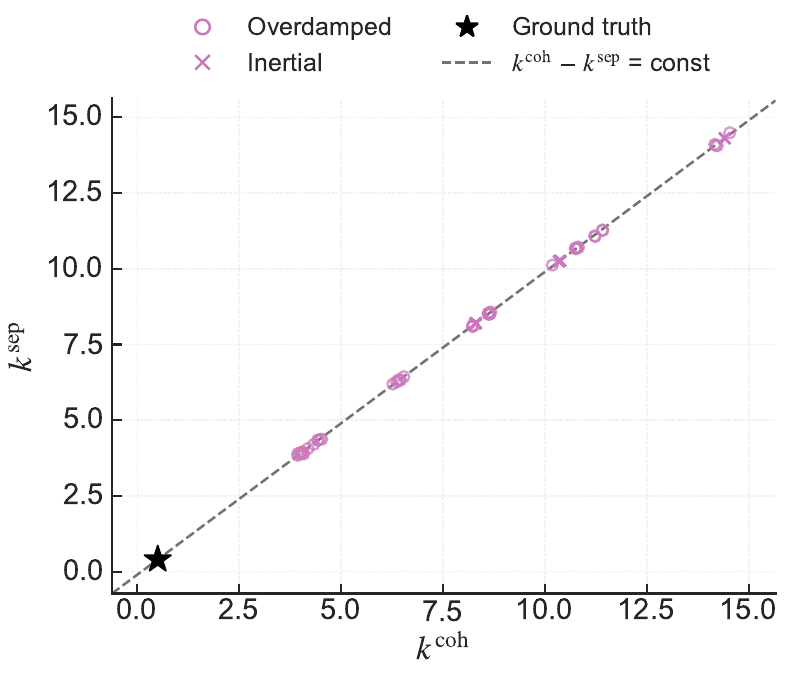}
            \phantomcaption\label{fig:degeneracy_scatter_TK2020}
        \end{subfigure}
        \hfill
        \begin{subfigure}[b]{0.48\textwidth}
            \textbf{(b)}\vspace{2pt}\\
            \includegraphics[width=\textwidth]{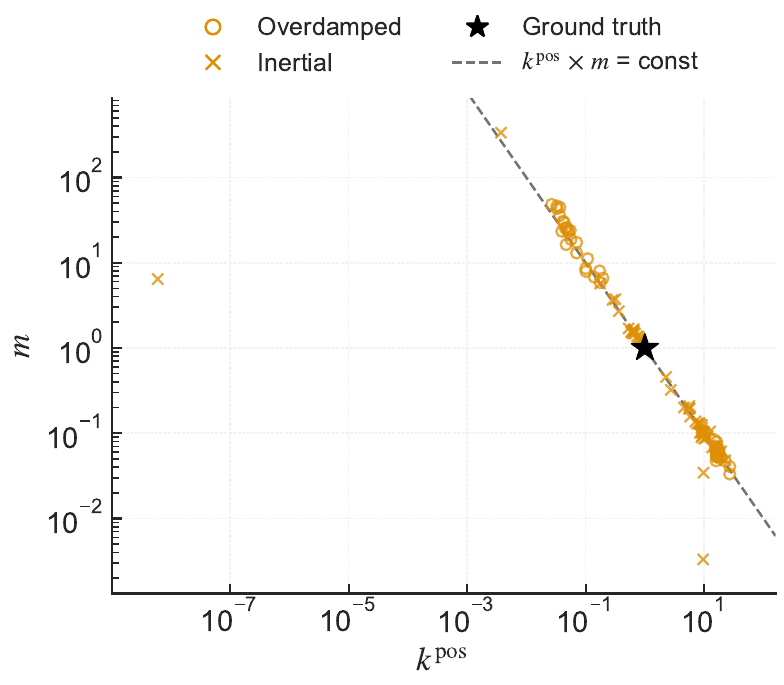}
            \phantomcaption\label{fig:degeneracy_scatter_KR2021}
        \end{subfigure}
        \\
        \begin{subfigure}[b]{0.60\textwidth}
            \textbf{(c)}\vspace{2pt}\\
            \includegraphics[width=\textwidth]{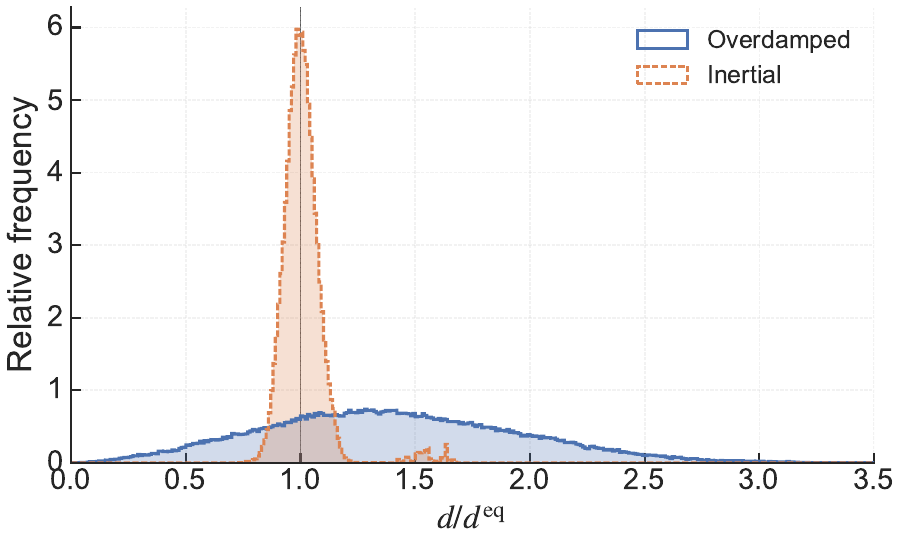}
            \phantomcaption\label{fig:relative_distance_kr2021}
        \end{subfigure}
        
        \caption{\textbf{(a), (b)}~Parameter estimates for TK2020 and KR2021, respectively (50 $\times$ 2 samples for each panel). The dashed lines indicate the theoretical degenerate manifolds, $k^{\mathrm{coh}}-k^{\mathrm{sep}}=\text{const}$ for TK2020 and $\hat{k}^{\mathrm{pos}}m=\text{const}$ for KR2021. The estimates are distributed along them in both cases.  
        \textbf{(c)}~Distribution of the mutual distance between neighbouring agents in the KR2021 models. For the Overdamped-KR2021 model, the distances fluctuate around $d^{\mathrm{eq}}$, with the peak of the distribution lying above it, while they are more sharply concentrated around $d^{\mathrm{eq}}$ for the Inertial-KR2021 model.}
        \label{fig:physical_origin}
    \end{figure}

    The simplest case is TK2020, whose social interaction becomes completely degenerate when the group consists of two agents. The social interaction term for agent 1 reads
    \begin{equation}
        \omega_1^{\mathrm{soc}} = \boldsymbol{e}^\perp_1\cdot \left[ (k^{\mathrm{coh}} - k^{\mathrm{sep}})\dfrac{\boldsymbol{x}_2-\boldsymbol{x}_1}{\|\boldsymbol{x}_2-\boldsymbol{x}_1\|} +k^{\mathrm{alg}}\dfrac{\boldsymbol{v}_2-\boldsymbol{v}_1}{\|\boldsymbol{v}_2-\boldsymbol{v}_1\|}\right],
    \end{equation}
    where $k^{\mathrm{sep}}$ and $k^{\mathrm{coh}}$ appear only through their difference $k^{\mathrm{coh}} - k^{\mathrm{sep}}$. Consequently, any perturbation along the line $k^{\mathrm{coh}} - k^{\mathrm{sep}} = \text{const}$ leaves the likelihood exactly invariant. Indeed, the estimates concentrate on this degenerate manifold (Figure~\ref{fig:degeneracy_scatter_TK2020}). As a result, estimation cannot single out the true parameter among this continuum of minima. This degeneracy is resolved when the group contains more than two agents (see Supplementary Information, Section 5.3 for details).

    The second case, KR2021, exhibits a milder form of degeneracy that emerges only near the ordered state. The KR2021 interaction switches from separation to cohesion at an equilibrium distance $d^{\mathrm{eq}}$, so the distance between neighbouring agents tends to approach this value (Figure~\ref{fig:relative_distance_kr2021}). Linearizing the interaction around $\|\boldsymbol{x}_j-\boldsymbol{x}_i\|\approx d^{\mathrm{eq}}$, we obtain
    \begin{equation}\label{eq:linearization_kr2021}
        \omega_i^{\mathrm{soc}} \approx \dfrac{1}{|\mathcal{N}_i|}\sum_{j\in \mathcal{N}_i}\left[ k^{\mathrm{alg}}\sin(\theta_j-\theta_i)+\hat{k}^{\mathrm{pos}}m(\|\boldsymbol{x}_j-\boldsymbol{x}_i\|- d^{\mathrm{eq}})\sin\phi_{ij} \right],
    \end{equation}
    where $\hat{k}^{\mathrm{pos}}$ and $m$ enter only through their product. The corresponding degenerate manifold is $\hat{k}^{\mathrm{pos}}m=\text{const}$, and the estimates are again distributed along it (Figure~\ref{fig:degeneracy_scatter_KR2021}). 

    Note that the validity of Equation~\eqref{eq:linearization_kr2021} does not require the mutual distance to converge to $d^{\mathrm{eq}}$. In the Overdamped-KR2021 model, the distance does not converge but fluctuates around $d^{\mathrm{eq}}$ (Figure~\ref{fig:relative_distance_kr2021}); however, these fluctuations are small compared with $1/m$ at the ground truth, so the argument of $\tanh$ remains within its linear range throughout the trajectory. The nonlinearity of the interaction, which is the only source of information that separates $\hat{k}^{\mathrm{pos}}$ from $m$, then enters only through the cubic term of the expansion, $-\hat{k}^{\mathrm{pos}}m^{3}(\|\boldsymbol{x}_j-\boldsymbol{x}_i\|-d^{\mathrm{eq}})^{3}/3$, and is therefore very weak. 
    
    This weak higher-order information partly explains the two distinct behaviours of the minimum eigenvalue. The samples in Figure~\ref{fig:envelope_kr2021_linear} correspond to the Overdamped-KR2021 estimates with $\hat m<m^*$ in Figure~\ref{fig:degeneracy_scatter_KR2021}, and those in Figure~\ref{fig:envelope_kr2021_fluctuation} to the estimates with $\hat m>m^*$. When $m$ is underestimated, Equation~\eqref{eq:linearization_kr2021} becomes even more accurate, and the interaction approaches complete degeneracy; accordingly, the minimum eigenvalue remains positive but grows only very slowly (Figure~\ref{fig:envelope_kr2021_linear}). When $m$ is overestimated, by contrast, the argument of $\tanh$ is no longer small, and the nonlinearity of the interaction should provide information that separates $\hat{k}^{\mathrm{pos}}$ from $m$. Nevertheless, the minimum eigenvalue fluctuates around zero within the observation time (Figure~\ref{fig:envelope_kr2021_fluctuation}). The origin of this behaviour remains unclear. One possible explanation is that the observed Hessian includes a residual-dependent term in addition to the information term, and this term may mask the slow growth of the latter when evaluated at an estimate away from the ground truth. Clarifying this point, for example by comparison with the expected Fisher information, is left for future work. 

    These two cases of TK2020 and KR2021 illustrate that unidentifiability can be either structural, embedded in the functional form of the model (TK2020), or dynamical, emerging from the state the system settles into (KR2021). While structural degeneracy requires reparameterization or reformulation of the model, dynamical degeneracy might be resolved by observing transient dynamics in which the deviation of the mutual distance from $d^{\mathrm{eq}}$ becomes comparable to $1/m$, a possibility that warrants further investigation.

\section{Discussion and Conclusion}
    \subsection{Summary of findings}
    As a first step towards the broader goal of promoting data-driven modelling in the study of flocking behaviour, we have proposed a diagnostic procedure to detect parameter unidentifiability in SDE models. This methodology helps identify the root cause of insufficient estimation precision, providing a guideline on whether to collect data over a longer period or to take other measures, such as reparameterization or reformulation of the model itself.

    The proposed methodology requires only a time-series plot of the minimum Hessian eigenvalue. Unidentifiable models are characterized by sublinear, stagnating fluctuations of this eigenvalue, while identifiable models exhibit linear growth in proportion to observation time. We also provided theoretical support for this procedure by deriving the asymptotic scaling of the minimum eigenvalue, utilizing the submartingale property and the Doob decomposition of the log-Hessian.
    
    The proposed diagnostic framework was applied to flocking models and revealed that TK2020 (for two-agent groups) is structurally unidentifiable and KR2021 practically unidentifiable. For the identifiable models, the worst relative error decayed in proportion to the inverse square root of the minimum eigenvalue, analogous to the classical relation between Fisher information and estimation variance. The growth rate of the minimum eigenvalue thus provides an approximate prediction of the observation length required to achieve a target precision. For the unidentifiable models, the root causes of unidentifiability were pinpointed as structural degeneracy inherent in the functional form and dynamical degeneracy that emerges around the ordered state, respectively. For TK2020, the numerical results were consistent with the theoretical $o(\sqrt{t}\log t)$ bound for completely degenerate models.
    
    \subsection{Limitations and future extensions}
    Despite the practical utility of the diagnostic procedure demonstrated in Section~\ref{sec:applications}, the methodology has several theoretical gaps and limitations. First, the theory guarantees the submartingale structure of the Hessian only at the ground truth $\boldsymbol{p}^*$, because it relies on the conditional expectation taken at $\boldsymbol{p}^*$. In practice, however, the Hessian is evaluated at the estimate $\hat{\boldsymbol{p}}$, and its robustness to this substitution, as well as its behaviour at local minima or saddle points away from $\boldsymbol{p}^*$, has not been established theoretically. Second, a mismatch between the generative model and the estimation model (model misspecification), which was absent in our simulations, may introduce additional difficulties. This issue is crucial when applying the method to empirical data, and its effect on the diagnostic remains to be investigated. Third, as the two distinct behaviours of the Overdamped-KR2021 model indicate, practically unidentifiable models may be difficult to distinguish from structurally unidentifiable ones. The scaling of the minimum-eigenvalue time series can depend on the estimated parameter values, which may make visual inspection misleading. Distinguishing between these two types of unidentifiability will require further refinement of the diagnostic, for example by removing the residual martingale term from the observed Hessian to isolate the systematic accumulation of information.
    
    In addition to these limitations, the methodology has a computational bottleneck: large memory consumption. Calculating the Hessian matrix via automatic differentiation requires a large amount of memory to retain the complex computational graph. While the models in this study have at most six parameters, some biological models involve hundreds~\cite{Hass2019}. Extending the method to such models will require more memory-efficient computation of the Hessian, which we leave for future work.

    Beyond these limitations, there are several pathways to broaden the applicability of this framework. One is to extend the procedure to multiple state variables per agent, allowing the diagnosis of flocking models in which the speed, assumed constant in this study, varies over time~\cite{DOrsogna2006, Sepulveda2013}. Adapting the framework to multivariable dynamics is straightforward provided that the noise terms are mutually independent Wiener processes, since the loss function then simply decomposes into a sum over variables. In contrast, incorporating cross-variable noise correlations or temporal dependencies would require further theoretical development. Another direction is to extend the procedure to non-Brownian noise processes, such as Poisson or L\'{e}vy processes. Such an extension would require replacing the Gaussian transition density (with separate treatment for infinite-variance processes), but would enable estimation and diagnosis for models with jumps, such as burst-and-coast models~\cite{Calovi2018, Li2021}, L\'{e}vy-flight foraging~\cite{Viswanathan1999}, and stochastic resetting~\cite{Evans2011}. 

    The methodology may also be applicable to ODE systems. Fitting deterministic models to long, noisy empirical data often suffers from severe trajectory divergence and rugged optimization landscapes~\cite{Ramsay2007}. By introducing small stochastic perturbations, deterministic models can be brought within the scope of our framework, in which case the diagnostic applies to the stochastic surrogate of the original ODE model. This approach not only makes the Hessian-based diagnosis applicable, but may also help smooth the likelihood geometry and quantify parameter sensitivities.
    
    \subsection{Comparison with existing methods and broader implications}
    Our results highlight two distinct types of ill-conditioned models: structurally unidentifiable models and practically unidentifiable models. While previous studies have characterized such ill-conditioning at a fixed data length~\cite{Gutenkunst2007, Transtrum2015}, the eigenvalue time series distinguishes the two types by their dynamical growth, and its slope quantifies how much data are required to reach a given precision.
    
    Traditionally, the distinction between structural and practical identifiability has been addressed using profile likelihood approaches~\cite{Raue2009}, which are well established in fields such as systems biology~\cite{Kreutz2013} and pharmacokinetics~\cite{Janzen2016}. While our method yields conceptually analogous conclusions regarding identifiability classifications, it offers complementary advantages. First, profile likelihood requires repeated nonlinear optimizations across parameter grids, making it computationally demanding for complex stochastic systems. In contrast, our approach requires only a single fitted estimate at the cost of evaluating the Hessian at multiple time points. Second, while profile likelihood assesses identifiability for a fixed dataset, our method tracks how the available information accumulates with observation time, thereby indicating whether a longer observation would improve precision. On the other hand, multi-grid profile likelihood provides the global shape of the degenerate manifold, while our method indicates the local tangent direction of the manifold at the estimate.
    
    Given the broader aim of advancing data-driven approaches in flocking studies, parameter estimation and diagnosis must ultimately be coupled with model selection methodology to identify the most suitable mechanism for empirical data. Standard model selection methods rely on information criteria such as the Akaike Information Criterion (AIC)~\cite{Akaike1974} or its adaptation to ergodic diffusion processes (Contrast-based Information Criterion, CIC)~\cite{Uchida2010}. However, the asymptotic theory underlying AIC and CIC fundamentally assumes the positive definiteness of the Fisher information matrix at the true parameter, which is violated by unidentifiable models. In such singular cases, alternative formulations such as the Widely Applicable Information Criterion (WAIC)~\cite{Watanabe2010} are theoretically desirable. In this context, our diagnostic framework serves as a useful preliminary check: linear growth, indicating non-degeneracy, supports the use of classical criteria (AIC/CIC), while sublinear stagnation signals the necessity of singular model selection techniques. 
    
    By revealing whether imprecise estimates call for more data or a different model, the proposed diagnostic turns identifiability from a hidden pitfall into a testable property, bringing mechanistic models of collective behaviour one step closer to the data.

\ethics{This research relies entirely on mathematical modelling and computer simulations. It does not involve human subjects, animals, or sensitive data, and therefore did not require ethical approval.}

\dataccess{The simulation data and results supporting this study are available on Zenodo at \url{https://doi.org/10.5281/zenodo.23535798}. The source code used to generate the results is also available on Zenodo at \url{https://doi.org/10.5281/zenodo.23535892}. Mathematical proofs, detailed experimental settings, and additional numerical results are provided in the Supplementary Information.}

\aucontribute{N.K. conceptualized the study, developed the mathematical framework, performed the computational simulations, and drafted the manuscript. H.K. supervised the project and revised the manuscript for important intellectual content. All authors gave final approval for publication and agree to be held accountable for the work performed therein.}

\competing{We declare we have no competing interests.}

\funding{This work was supported by JSPS KAKENHI Grant Number JP25K15258 and the JSPS Core-to-Core Program ``Advanced core-to-core network for the physics of self-organizing active matter'' (JPJSCCA20230002).}

\ack{We thank Dr Yusuke Kato (University of Michigan) for fruitful discussions, particularly regarding the estimation method. We also thank Dr Takeru Matsuda (The University of Tokyo) for helpful discussions and valuable statistical advice following a presentation of this work.}

% \disclaimer{Insert disclaimer text here.}

%%%%%%%%%% Insert bibliography here %%%%%%%%%%%%%%

\printbibliography

\end{document}